\documentclass[preprint,showpacs,preprintnumbers,amsmath,amssymb,superscriptaddress]{revtex4}

\usepackage{graphicx,graphics}   

\usepackage{graphicx,amsfonts}
\usepackage{epsfig}

\usepackage{dcolumn}
\usepackage{bm}
\begin{document}

\title{The angular momentum of two massless fields revisited }

\author{V. Pleitez}%
\email{vicente@ift.unesp.br}
\affiliation{
Instituto  de F\'\i sica Te\'orica--Universidade Estadual Paulista \\
R. Dr. Bento Teobaldo Ferraz 271, Barra Funda\\ S\~ao Paulo - SP, 01140-070,
Brazil
}

\date{08/31/15}
%
\begin{abstract}
We consider the angular momentum of two massless fields using the Landau's arguments. In particular, we point out the explicit and implicit assumptions made by Landau to obtain the proof that a spin one system cannot decay into two photons. 
We show also under what conditions the result can be avoided. 
\end{abstract}

\pacs{ 13.40.Hq 
14.70.Bh 
11.30.Er 	
}     

\maketitle

\section{Introduction}
\label{sec:intro}

That a stationary system with spin-1 cannot decay into two photons was first observed  
by Landau~\cite{landau} and then, independently, by Yang~\cite{yang}. This is now called the Landau-Yang theorem and it  was used to ruled out the possibility that the resonance with a mass of near 125 GeV, decaying into two  photons, is a spin-1 particle and, for this reason only the possibilities of spin-0 and 2 remained~\cite{Aad:2012tfa,Chatrchyan:2012ufa}. However, later on, evidence that its spin is really zero and the parity even was found also by the CERN collaborations~\cite{Chatrchyan:2012jja,Aad:2013xqa}.   
Anyway this brought back to the attention of the elementary particle physics community the theorem. Notwithstanding, only the consequences of the theorem are used without enter into details. Here we revisit the theorem and emphasize under what conditions it is valid, and  also point out the consequences if some of these conditions are given up. We also consider the case of two massless (pseudo)scalars and fermions and two gluons. In the latter case we show that the theorem is evaded. Although we follow Landau's, Yangs's arguments are briefly commented too. 

The outline of this work is the following. In Sec.~\ref{sec:lyt} we discuss the assumptions under which the theorem is valid. In Sec.~\ref{sec:2photons} we reproduce the Landau's and Yang's arguments for the case of the two photons system. In Sec.~\ref{sec:2s} we study the case of two massless scalars or/and pseudo-scalars. The case of two massless spin-1/2 fermions is consider in Sec.~\ref{sec:2fermions}. In Sec.~\ref{sec:failure} we consider how the theorem is evaded if some of the basic assumptions considered in Sec.~\ref{sec:lyt} are not used. However, in the case of two gluons which is consider in Sec.~\ref{sec:2gluons} the theorem is always evaded.
The last section is devoted to the conclusions and some speculations about the evasion of the theorem.

\section{The basic assumptions of the Landau-Yang theorem}
\label{sec:lyt}

There are explicit and implicit assumptions needed to show the theorem. 
The explicit assumptions are: 
\begin{enumerate}
\item[\textbf{i)}] Although the separation of the total angular momentum into the spin and orbital angular momentum $J=S+L$ has no physical meaning for massless particles, Landau used it just as a mathematical trick 
(but it is justified by the fact that the angular momentum of the two photon system 
can have only values that are consistent with those of the decaying system which is a massive one, and in this case that decomposition is valid); 
\item[\textbf{ii)}] The spin of the two photons is determined by the range of the tensor (scalar, vector, ...) and the orbital angular momentum by the order of the spherical harmonics describing the states. 
\item[\textbf{iii)}] The parity of the states is determined by how the respective tensor behaves under $\hat{n}\to -\hat{n}$, where $\hat{n}$ is the unit vector directed from one photon to the other. The angular momentum, $L$, is chosen in order to keep the parity obtained in this way. 
\item[\textbf{iv)}] Bose symmetry. This is well motivated since the symmetry under permutation of the fields 
defines when a system is a boson i.e., if it obeys the Bose-Einstein statistics. 
\item[\textbf{v)}] The photons are in the center of momentum frame i.e., $\vec{p}_1+\vec{p}_2=0$. This condition is not relevant since there exist inertial observers for which $\vec{p}_1+\vec{p}_2\not=0$ and the result must continued to be valid. However, the calculations are easier in this frame.
\item[\textbf{vi)}] Rotational invariance (Lorentz invariance). It is also well motivated since this implies the conservation of the angular momentum. 
\item[\textbf{vii)}] The photons have to be real (transverse). 
\end{enumerate}

There are also implicit assumptions:
\begin{enumerate} 
\item[\textbf{a)}] Parity, $P$, is conserved. Photons can be in two helicity states with the same strength in each of them, and the decaying system is a parity eigenstate with values of $J^P$  well defined.  
\item[\textbf{[b]}] The $CPT$ symmetry is conserved.
\item[\textbf{c)}] The spacial dimensions are three. 
\item[\textbf{d)}] The space-time dimensions are commutative (consequence of the Lorentz invariance).
\end{enumerate} 

Of these conditions only iv), vi), vii), [c] and [d] allow the theorem to be evaded. 

\section{The angular momentum of two photons}
\label{sec:2photons}

The wave function of the system formed by two photons is represented as a second rank tensor $E_{ik}$ including the angular momentum part, written in terms of spherical harmonics. This tensor is  build with each photon 
component being linear in each of them. Under the assumption of rotational invariance, this tensor
is a function only of $\hat{n}$, defined as  $\vec{r}_2-\vec{r}_1=\hat{n}r$, and $r$ is fixed. 

The photons are assumed to be real (on-shell) by imposing the transversal conditions:
\begin{equation}
E_{ik}n_k=0,\quad E_{ik}n_i=0.
\label{real}
\end{equation}  
Moreover, the permutation of the photon indices in the tensor, and also making $\hat{n}\to -\hat{n}$, 
must be even since they are bosons
\begin{equation}
E_{ik}(\hat{n})=E_{ki}(-\hat{n}).
\label{bose}
\end{equation}
It is always possible to write a second rank tensor $E_{ik}$ as the sum of a symmetric ($S_{ik}$) and an antisymmetric
($A_{ik}$) tensors
\begin{equation}
E_{ik}=S_{ik}+A_{ik}.
\label{sum}
\end{equation}

The relations (\ref{real}) and (\ref{bose}) must be independently satisfied by each tensor in (\ref{sum}), they have to transform under the permutation of the indices and $\hat{n}\to -\hat{n}$, hence:
\begin{equation}
S_{ik}(\hat{n})=S_{ik}(-\hat{n}),\quad A_{ik}(\hat{n})=-A_{ki}(-\hat{n}).
\label{bose2}
\end{equation}

These conditions determine the parity of the wave functions. In this referential frame (the sum of the momenta 
of the photons is zero)  the  parity transformation is equivalent to $\hat{n}\to -\hat{n}$. Thus, under a parity transformation, the states related to $S_{ik}$ are even and those related to $A_{ik}$ are odd. The conservation of parity implies that the two photon system is even or odd under parity and not a linear combination of both sort of states and also that the photons can be in two possible states of helicity, say left- and right-handed polarized.
  
Let us first consider the states that can be built from the antisymmetric tensor $A_{ik}$. 
If the spacial dimensions are three, as it was implicitly assumed, an antisymmetric tensor $A_{ik}$ is dual to a vector $A_l$, and we can write $A_l=\varepsilon_{lik}E_{ik}$. 
Then, the conditions $E_{ik}n_k=(1/2)\varepsilon_{ikl} A_ln_k$ is just a consequences of the number of spacial dimensions and does not depend on the transversality conditions. However, since we are assuming the conditions in Eq.~(\ref{real}), it implies $\vec{A}\times \hat{n}=0$: The vector $\vec{A}$ is proportional to $\hat{n}$ 
and we can write $\vec{A}(\hat{n})=\phi(\hat{n})\hat{n}$, where $\phi(\hat{n})$ is a scalar function $\phi(\hat{n})=\phi(-\hat{n})$. Hence, the vector character of $A_i$ explains its odd parity and the true tensor character is given by the scalar $\phi(\hat{n})$ that is, that the spin of the system is $S=0$. Since these states are described by a scalar function, $\phi(\hat{n})$, they have to be constructed from spherical harmonics of even order, zero included. Moreover, since $J=L$ with $L$ even are allowed, we obtain $J^P=0^-$ and $J^P=(2k)^-$ as the only possibilities.
 Summarizing, in the case of the states that can be built from the antisymmetric tensor $A_{ik}$, it is the number of spacial dimensions (three) \textit{plus} the transversality of the photons, that do not allow states with $J^P=1^-,(2k+1)^-$, $k\geq1$. 

Next, let us consider the symmetric tensor $S_{ik}$. This tensor can be decomposed in two parts as follows: $S_{ik}~=~\delta_{ik}\textrm{Tr}\,S~+~S^\prime_{ik}$ with $\textrm{Tr}\ S^\prime_{ik}=0$. Since both of them are even under parity, they are expressed in terms of spherical harmonics of even $L$. Since $\textrm{Tr}\, S$ is a scalar, it corresponds to $S=0$  then $J=L$ with $L$ an even integer (zero included). Then, there are one state $J^P=0^+$ and one $J^P=(2k)^+$ states (for a given $J$).  The traceless $S^\prime$ correspond to spin $S=2$ and we have that ($L$ is even): i) for even $J\not=0$ there are three possible states $L=J-2,J,J+2$, i.e., $nJ^P=3(2k)^+$, and for odd $J\not=1$ there are two states $L=J-1,J+1$, i.e., $nJ^P=2(2k+1)^+$. When 
$J=0$ and $J=1$ there is in each case one states with  $L=2$, $J^P=0^+$ and $J^P=1^+$, respectively. Thus, at first sight, from $S_{ik}$, the allowed wave functions are those with: 
\begin{equation} 
(a)\;\; \textrm{Tr}\,S\,: \;J^P=0^+, J^P=(2k)^+;\;\; 
(b)\;\;S^\prime_{ik}\,:nJ^P=0^+,1^+,3(2k)^+,2(2k+1)^+. 
\label{symmetric}
\end{equation}

However, because of the condition in Eq.~(\ref{real}), we have to take into account again the orthogonality between $S_{ik}$ and $\hat{n}$. It means that we have to exclude from those states in (\ref{symmetric})b, those states corresponding to a symmetric second rank tensor parallel to the vector $\hat{n}$. These sort of tensors are of the form $B_in_k+B_kn_i$, with $B_i(\hat{n})$ being a vector (change sign under $\hat{n}\to-\hat{n}$). This vector $\vec{B}$ correspond to spin $S=1$ and odd $L$ in order to obtain states with even parity as those from the $S^\prime_{ij}$. For every even $J\not=0$, there are two possible states $L=J-1,J+1$, i.e., there are two states $nJ^P=2(2k)^+$. For each odd $J$  there is one state with $J=L$, it means states $J^P=1^+,(2k+1)^+$. When $J=0$ we have only one state with $L=1$, and  $J^P=0^+$. 
Hence, from the $S_{ik}$ states in (\ref{symmetric}) we have to eliminate, in order to have transverse photons, the following states: 
\begin{equation} 
n(J)^P=0^+,1^+,2(2k)^+,(2k+1)^+.
\label{dois}
\end{equation} 
The allowed states from $S_{ij}$ are the difference between those in Eq.~(\ref{symmetric}) and those in Eq.~(\ref{dois}).The condition in Eq.~(\ref{real}) eliminates some of the states that would be allowed for off-shell photons. The states of angular momentum allowed for two photons are summarized in the first line of the Table. 

Notice that the Landau's argument has nothing to do with the existence or not of a direct interactions of the two photons with the decaying system. In fact, in the known cases the connection is through loop effects: scalar $h\to \gamma\gamma$ or an pseudo-scalar $\pi^0\to\gamma\gamma$.

Yang use the following argument to ruled out the possibility that a vector or pseudo-vector  
can decay into two real photons: these photons has four states $E_{RR}, E_{LL}, E_{LR}$ and $E_{RL}$, where $R$ and $L$ denote the circular polarization of each states. The 
combination $E_{RR}+E_{LL}$ and $E_{RR}-E_{LL}$ have eigenvalues $+1$ and $-1$ under parity, respectively, but the four states have the eigenvalue $+1$ under rotation of $\pi$ around the $x$-axis. For an odd initial state the only possibility is to decay to $E_{RR}-E_{LL}$. For an even initial state there are three possibilities: $E_{RL},E_{LR}$ and $E_{RR}+E_{LL}$. Notwithstanding, under a rotation of $\pi$ around the $x$-axis the decaying system behaves as $(-1)^J$, and $J=1$ the photon change sign but the two photon do 
not: or parity or angular momentum are conserved but not both at the same time. This forbids the decaying system to be an spin one system. But the initial state has to have even intrinsic parity. If it has no definite parity it has one component of odd parity and the behavior under the $\mathcal{R}_z(\pi)$ is as $-J^P$, then in this case the system can decay to the  $E_{RR}-E_{LL}$ two photon state. This argument is the one most used in literature, to show that the neutral pion has spin-0, see for instance Ref.~\cite{kallen}.

\section{Two massless scalar or pseudo-scalar bosons }
\label{sec:2s}

Let us consider the case of two  massless spin-$0$ bosons. They can be two scalars, $\phi^0_s$, two pseudo-scalar, $\phi^0_p$, or  one scalar and one pseudo-scalar. 
We work, as in the previous case, in the $\vec{p}_1+\vec{p}_2=0$ frame. 
In this case the tensor is always a scalar or a pseudo-scalar. 
Denoting $\Phi^0(\hat{n})$ the system of two scalars $\phi^0_s\phi^0_s$ or two pseudo-scalars $\phi^0_p\phi^0_p$, the system has the angular momentum  determined by the order of the spherical harmonic and since $\Phi(\hat{n})$ is a scalar i.e., $\Phi^0(\hat{n})=\Phi^0(-\hat{n})$, the system has even parity, spin-0, and thus $L$ even.  Hence we have always $J=L$ and the states are $J^P=0^+,(2k)^+$. 
Next, let us consider a system with one scalar and one pseudo-scalar massless fields denoted $\Psi^0\equiv\phi^0_s\phi^0_p$. The system satisfies the condition $\Psi^0(\vec{n})=-\Psi^0(-\vec{n})$ the respective spherical harmonic are of order odd and the parity is odd also, hence $L$ is odd.  Again, $J=L$ and the states allowed have $J^P=1^-,(2k+1)^-$.
We note that the decays into $0^-,1^\pm,(2k)^-$ are forbidden when both fields are of the same type two scalars or two pseudoscalars, while the forbidden states are $0^\pm,1^+,(2k+1)^+$ when one of the field is an scalar and the other a pseudo-scalar. Both cases, $\Phi^0$ and $\Psi^0$, are summarized in the second and third lines in the Table, respectively.

\section{Two massless spin-1/2 fermions}
\label{sec:2fermions}

Let us consider a system of two massless fermion $F^0\equiv f^0f^0$ which do not carry any quantum number i.e., they are truly neutral particles.
In this case the wave function of the system is antisymmetric under the exchange of the fermions and $\hat{n}\to-\hat{n}$. Hence, the system is defined by a second rank tensor $\mathcal{F}_{ik}$ such that 
\begin{equation}
\mathcal{F}(\vec{n})_{ik}=-\mathcal{F}_{ki}(-\vec{n}),
\label{ff}
\end{equation}
and the states from $\mathcal{F}_{ik}$ have the decomposition into symmetric and antisymmetric parts:
\begin{equation}
\mathcal{F}_{ik}=\mathcal{F_S}_{ik}+\mathcal{F_A}_{ik},
\label{ff2}
\end{equation}
but now with 
\begin{equation}
\mathcal{F_S}_{ik}(\vec{n})=-\mathcal{F_S}_{ik}(-\vec{n}),\quad \mathcal{F_A}_{ik}(\vec{n})=\mathcal{F_A}_{ki}(-\vec{n}).
\label{ff3}
\end{equation}
Notice that now, under a parity transformation the states related to $\mathcal{F_S}_{ik}$ are odd  and those related to $\mathcal{F_A}_{ik}$ are even. However, in this case, the indices $i,j$ run over a 2-dimensional space and there is no constrains like those in Eq.~ (\ref{real}). 
The spin of the states that we can build from $\mathcal{F_A}_{ik}$ has $S=0$ then $J=L$ with $L$ odd.  Moreover, since in a two-dimensional space there is only one antisymmetric tensor $\epsilon_{ij}$, hence $\mathcal{F_A}_{ik}=\epsilon_{ik}\chi(\hat{n})$, and $\chi(\hat{n})$ is a pseudoscalar function $\chi(-\hat{n})= -\chi(\hat{n})$, it means that it is expanded in spherical harmonics with $L$ odd. The states allowed are $1^+,(2k+1)^+$ and, as before, $k\geq1$. 

The states from the symmetric tensor have $S=1$ and odd parity, then $L$ is odd. 
Hence, for even $J\not=0$ we have two states $J-1,J+1$, i.e., $nJ^P=2(2k)^-$ and those with odd $J$ have $J=L$ have $J^P=1^-,(2k+1)^-$, finally for $J=0$ we have one state with $L=1$ and $J^P=0^-$. 
Summarizing, we have the allowed states for two massless fermions are: $0^-,1^\pm,2(2k)^-,(2k+1)^\pm$. 
Notice that systems with $J^P=0^+,(2k)^-$ cannot decay into two massless fermions. 

It is important to note that, in order to obtain these results we have assumed that the fermions can have both helicity states i.e., that they are vectorial fermions (parity is conserved). The result is different for chiral fermions. For instance, if the fermions can be only in one helicity state, say left-handed,  only the states with $S=0$ can occurs, i.e., $J^P=1^+,(2k+1)^+$.
In this case, the decay $0^-\to ff$ is possible only if the fermions are massive, and one of them is in the "wrong" helicity. For instance, $K_L\to\nu\bar{\nu}$ may occur only if neutrinos are massive~\cite{Marciano:1996wy,Gninenko:2015mea}. 

\section{Avoiding the theorem in the two photons system}
\label{sec:failure}

In this section we considered some situations in which the Landau-Yang theorem can be avoided. Some of the assumptions are trivial in the sense that their violation avoid the result of the theorem. For instance, the invariance under rotations. In this case the angular momentum is not conserved an any thing is allowed. Here, we will concerned only on those assumptions that if given up, some interesting consequences may happen.   

\subsection{Wrong Bose symmetry}
\label{subsec:bose}

Returning to the two photons decay we now assume the extreme unrealistic case when photons do not satisfy the Bose symmetry: their wave function change sign when they are permuted and together $\hat{n}\to-\hat{n}$. Hence, the system is defined by the second rank tensor $\mathcal{O}_{ik}$ such that 
\begin{equation}
\mathcal{O}(\vec{n})_{ik}=-\mathcal{O}_{ki}(-\vec{n}),
\label{odd}
\end{equation}
and the states from $\mathcal{O}_{ik}$ have the usual splitting into a symmetric and an antisymmetric parts:
\begin{equation}
\mathcal{O}_{ik}=\mathcal{S}_{ik}+\mathcal{A}_{ik},
\label{sum2}
\end{equation}
but now with 
\begin{equation}
\mathcal{S}_{ik}(\vec{n})=-\mathcal{S}_{ik}(-\vec{n}),\quad \mathcal{A}_{ik}(\vec{n})=\mathcal{A}_{ki}(-\vec{n}).
\label{odd2}
\end{equation}
Thus, under a parity transformation, the states related to $\mathcal{S}_{ik}$ are odd  and those related to $A_{ik}$ are even. The case appears at first sight as the case of two fermions in Eq.~(\ref{ff}). However, now we have to take into account the transversality conditions (\ref{real}).

By a similar argument as in Sec.~\ref{sec:2photons}, we can show that the states from the tensor $\mathcal{A}_{ij}$ has even parity and that this tensor is now  equivalent to a pseudo-scalar function ~$\phi(\hat{n})=-\phi(-\hat{n})$. Hence, the states have $S=0$ and with the appropriate parity have to be constructed from spherical harmonic of odd order. Then, $J=L$ and the wave functions have $J^P=1^+$ and $J^P=(2k+1)^+$. We see that the theorem is evaded in the states obtained with $\mathcal{A}_{ij}$. 

Next, let us consider the tensor $\mathcal{S}_{ij}$ whose states have odd parity. This parity is implemented by chosen $L$ odd. As before, we use the decomposition $\mathcal{S}_{ik}=\delta_{ik}\textrm{Tr}\,\mathcal{S}+\mathcal{S}^\prime_{ik}$ with $\textrm{Tr}\,\mathcal{S}^\prime=0$. The states from $\textrm{Tr}\,\mathcal{S}$ have $S=0$ and $J=L$ but since now we have a pseudo-scalar function ($\mathcal{O}$ is a pseudo-tensor), they are expanded using spherical harmonics of odd order, then $J$ is odd and the allowed wave functions can have $J^P=1^-,(2k+1)^-$. 

The states from the traceless $\mathcal{S}^\prime$ correspond to spin $S=1$ 
and for odd $L$  we have that: i) for a given even $J\not=0$ there are two possible states $L=J-1,J+1$ i.e., 
$nJ^P=2(2k)^-$, and for odd $J$ the  states have $J=L$, i.e., $nJ^P=1^-,(2k+1)^-$. When 
$J=0$ there is one states with  $L=2$ and $J^P=0^-$. Thus, from 
$\mathcal{S}_{ik}$ the wave functions are: $J^P=1^-,(2k+1)^-$ from $\textrm{Tr}\,S$, and 
$J^P=0^-,1^-,2(2k)^-,(2k+1)^-$ from $\mathcal{S}^\prime_{ik}$.
   
However, because the condition in Eq,~(\ref{real}), we have to subtract from $\mathcal{S}$, the states obtained from second rank tensor that are parallel to $\hat{n}$ with odd parity. They are of the form $V_in_k-V_kn_i$.  Since we have $V_i(\vec{n})=+V_i(-\vec{n})$ the states built from this pseudo-vector have $S=1$ and odd parity. Hence, we have expand the vector $V_i$ in term of spherical harmonics of odd order.
For even $J\not=0$ we have two states $J-1,J+1$, i.e., $nJ^P=2(2k)^-$; for odd $J$ we have $J=L$ and the states are $J^P=1^-,(2k+1)^-$; and, finally, when $J=0$ we have one state: $J^P=0^-$. After subtracting the states from $V_i$ we have that the states allowed are only $1^\pm,(2k+1)^+$. We see that the states with $J^P=1^\pm$ can decay into two photons and the theorem is evaded. This unrealistic situation was chosen just to show the importance of the bosonic nature of the photons in the proof of the model. However, see Sec.~\ref{sec:con}.

\subsection{Off-shell photons}
\label{sec:off}

One crucial argument in Sec.~\ref{sec:2photons} is the fact that the two photons are on-shell and they have to satisfy the conditions in (\ref{real}).  If we given up this condition the tensor $A_{ij}$ still has a dual vector but now $\vec{A}\times \hat{n}\not=0$, then the vector $\vec{A}$ is not proportional to $\hat{n}$.   The relation $A_l=\phi(\hat{n})\hat{n}$ follows only because of the conditions (\ref{real}) are also assumed. If this were not the case the tensor $A_{ij}$ would admit spin-1 states. The case is more evident for the symmetric wave functions, $S^\prime_{ij}$. The states that are orthogonal to this tensor, $B_jn_k+n_jB_k$ were subtracted from those of $S^\prime_{ij}$ again because the condition (\ref{real}). In this case the states allowed are those in (\ref{symmetric}). It means that a system of spin-1 can decay into two off-shell photons when (\ref{real}) is not valid. In this case all the states in (\ref{symmetric})b are allowed. We see that a spin-1 states cn decay into two photons if the later are off-shell.

\section{QCD: Two gluons system}
\label{sec:2gluons}

In this case we have to consider also the color degrees of freedom. In this case 
the two gluons system is represented also by a second rank tensor $E^{ab}_{ik}(\hat{n})$ and the transversality condition reads in this case
\begin{equation}
E^{ab}_{ik}n_k=E^{ab}_{ik}n_i=0.
\label{real2}
\end{equation}
The permutation of the gluons indices because of the Bose symmetry implies
\begin{equation}
E^{ab}_{ik}(\hat{n})=E^{ba}_{ki}(-\hat{n}).
\label{bose3}
\end{equation}

As in Sec.~\ref{sec:2photons}, the second rank tensor can be written as the sum of a symmetric 
\begin{equation} 
E^{ab}_{ik}(\hat{n})=S^{ab}_{ik}+A^{ab}_{ik}.
\label{sum3}
\end{equation}
but now
\begin{equation}
A^{ab}_{ik}=A^{\{ab\}}_{[ik]}(\hat{n})+A^{[ab]}_{\{ik\}} (\hat{n}),\quad
S^{ab}_{ik}=S^{\{ab\}}_{\{ik\}}(\hat{n})+S^{[ab]}_{[ik]} (\hat{n}),
\label{sum4}
\end{equation}
where as usual $\{ab\}$ and $[ab]$ denote symmetric and antisymmetric indices, respectively; the same notation is used in the indices $i,k$. 
The relation  (\ref{bose3}) must be independently satisfied by each tensor in (\ref{sum3}), hence
\begin{eqnarray}
&& S^{\{ab\}}_{\{ik\}}(\hat{n})=S^{\{ab\}}_{\{ik\}}(-\hat{n}),\quad S^{[ab]}_{[ik]}(\hat{n})=S^{[ba]}_{[ki]}(-\hat{n}),\nonumber \\ && A^{\{ab\}}_{[ik]}(\hat{n})=-A^{\{ba\}}_{[ki]}(-\hat{n}),\quad
A^{[ab]}_{\{ik\}}(\hat{n})=-A^{[ba]}_{\{ki\}}(-\hat{n}),
\label{bose2}
\end{eqnarray}
as also the transversality conditions in (\ref{bose3}) have to be satisfied by each tensor in (\ref{sum4}). From (\ref{sum4}) we note that the parity of $S^{ab}_{ik}$ is even and that of $A^{ab}_{ik}$ odd. 
Let us consider first the states related with the odd parity tensor $A^{ab}_{ik}$.
The states built form $A^{[ab]}_{\{ik\}}$ are as those in the two photons system in Sec.~\ref{sec:2photons}: they have $S=0$ and odd parity and, by the same arguments for the photons, even order $L$. Hence the theorem is valid for this tensor. The states from $A^{\{ab\}}_{[ik]}$ have also $S=0$ but now odd parity, and odd order $L$, as in the case of Sec.~\ref{subsec:bose} the state $J^P=1^+$ is allowed and the theorem is evaded.
Next, let us consider the states from the tensors with even parity related with the tensor $S^{ab}_{ik}$. The tensor  $S^{\{ab\}}_{\{ik\}}$ corresponds to the same states as the $S_{ik}$ in the two photons system after excluding the states corresponding to the tensor parallel to the vector $\hat{n}$, which can be written down in the form $B^{\{ab\}}_in_k+B^{\{ab\}}_kn_i$. After doing this the state $J^P=1^+$ is forbidden. Finally, the tensor $S^{[ab]}_{[ik]}$ correspond the states with odd order $L$ and we have to exclude the states corresponding the the tensor parallel to $\hat{n}$ written down in this case as $V^{[ab]}_in_k-V^{[ab]}_kn_i$. As in Sec.~\ref{subsec:bose}, the state with $J^P=1^-$ is permitted. The theorem is violated again. Summaryzing, in QCD the Landau-Yang theorem is evaded even the two gluons system obeying the Bose symmetry, since the tensors $A^{\{ab\}}_{[ik]}$ and $S^{[ab]}_{[ik]}$ allow states with $J^P=1^\pm$. The case of $S^{[ab]}_{[ik]}$ was first noted in Ref.~\cite{Beenakker:2015mra}.

\subsection{Extra dimensions}
\label{subsec:extrad}

In the previous discussions it has been assumed that there are only three spacial and one time dimensions. 
This was used in order to use the existence of a dual vector of the antisymmetric tensor. If extra spacial dimensions do exist think can be different. A detailed study of this case is beyond the scope of this paper. However, we call the attention that
in $D+1$ space-time dimensions, the little group leaving invariant a massless (massive at rest) particle  is $O(D-2,1)$ $(O(D-1,1))$~\cite{Weinberg:1984vb}. In this cases the transversality will be only approximated in our three spacial dimensions and  for this reason the theorem also will be evaded. In this case also it is possible to have Lorentz invariance violations and the angular momentum in our three spacial dimensions could be not conserved. All these situations will imply violations of the Landau-Yang' theorem. 

\subsection{Discrete symmetries}
\label{subsec:discrete}

We note that parity ($P$), charge conjugation ($C$), and time reserval ($T$) violations do not imply evoiding the theorem only the features of the two photons. For instance, orto-positronium $^1S_0$ with $S=0,m_s=0$ (singlet state) can decay into two photons because the $C$ conservation \textit{and} the Landau-Yang theorem; while para-positronium $^3S_1$ with $S=1$ and $m_s=-1,0,+1$ (triplet state) can only decay into three photons by $C$ conservation and the Landau-Yang theorem. However, even if $C$ invariance were violated the theorem forbids the $^3S_1\to\gamma\gamma$ decay. The same happens by the $T$ violation and a possible $CPT$ violation. We recall that $CPT$ invariance and the spin-statistic theorem are no necessarily connected. For instance, if parity were violated, say by existing only photons of a given helicity, according to the analysis in Sec.~\ref{sec:2photons} only the states from the antisymmetric tensor $A_{ij}$, which have $S=0$ can be built, but the transversality conditions still eliminate the $J=1$ state and the theorem is not violated even in this unphysical situation.    

\section{Conclusions}
\label{sec:con}

We would like to stress that under the general, explicit and implicit, conditions considered in Sec.~\ref{sec:intro} are satisfied, even if photons obey Bose symmetry, as in fact they do, it is the on-shell conditions plus the tree spacial dimensions that  eliminates the $J=1^\pm$ possibility for the two (real) photons system. For virtual photons $1^-\to \gamma^*\gamma^*$ through the $A_{ij}$ interactions and $1^+\to \gamma^*\gamma^*$ through the $S^\prime_{ij}$ interactions are allowed. Also, if the decaying system is off-shell, the decays $J^P=1^\pm\to \gamma\gamma$ are also allowed. In this case the decaying system has not a well defined mass. 

Since a virtual spin-1 can decay into two real photons, not any evidence of this sort of decay may be an indication of new physics (NP). Hence, $Z\to NP\to \gamma\gamma$ has, if it does exist, to be larger than $Z^*\to\gamma\gamma$ in the SM. We would like to mention that experimentally an upper limit for the decay $Z ~\to~\gamma\gamma,\pi^0\pi^0$  has been obtained recently at CDF:  $Br(Z\to \gamma\gamma)<1.46\times10^{-5}$ and $Br(Z\to \pi^0\pi^0)<1.52\times10^{-5}$~\cite{Aaltonen:2013mfa}. These are not too small numbers. It means that, if new physics, for instance with violation of the Lorentz invariance~\cite{Behr:2002wx}, may induces such decays, since in the SM the decay $Z^*\to \gamma\gamma$ is allowed, it is certainly a background for the new physics violating the Landau-Yang theorem. \cite{Gninenko:2011ws}

It is important to realize that the two photons carry information of the decaying system and that we do not know \textit {a priori} all sort of systems that could decay in this way.
We can unified the cases Secs.~\ref{sec:2photons} and \ref{subsec:bose} and define a tensor ($\mathcal{E}_{ik}$) and an odd tensor ($\mathcal{O}_{ik}$) in the following way: 
\begin{equation}
E_{ik}(\vec{n})=a\,\mathcal{E}_{ik}(\vec{n})+b\,\mathcal{O}_{ik}(\vec{n}),
\label{pnook}
\end{equation}
with 
\begin{equation}
\mathcal{E}_{ik}(\vec{n})=E_{ki}(-\vec{n}),\quad \mathcal{O}_{ik}=-\mathcal{O}_{ki}(-\vec{n}),
\label{pok}
\end{equation}
where$a$ and $b$ are dimensionless parameters. 
When $a=1$ and $b=0$ we have the Landau's case i.e., Bose' photons. 
In general $\vert a\vert^2+\vert b \vert^2=1$ but, if $b\not=0$, it must satisfy  $\vert b\vert \ll \vert a\vert$. 

An interesting feature is that the analysis of the angular momentum of the two photons system may be used to search for exotic matter with opposite parity. For instance, if $\pi^0$ partner with $J^P=0^+$ does exist~\cite{Foot:2014mia}, the physical pion must be a linear combination of both $0^-$ and $0^+$ fields and this should appear in the angular distribution of the two photons. We can interpret the old result by Plano et al.,~\cite{plano} as having shown that, if the neutral pion has a component of opposite parity, its amplitude has to be much smaller than the odd parity amplitude, but it does not need to be zero. Although the violation of the Bose symmetry is not possible within the context of local quantum field (QFT)~\cite{Sahoo:2014nna}, if some deviation of this symmetry is observed it will certainly imply new theoretical framework beyond QFT and the smoking gun may be the violation of the Landau-Yang theorem.


\newpage

\begin{table}[ht]
\begin{tabular}{|l||c| }\hline  
& $J^P$   \\ \hline \hline
$\gamma\gamma$ & $0^-,(2k)^-$; $0^+$, 2$(2k)^+$,$(2k+1)^+$  \\ \hline
$\Phi^0$ & $0^+,(2k)^+$ \\ \hline
$\Psi^0$ & $1^-,(2k+1)^-$ \\ \hline
$F^0$ & $0^-,1^-,2(2k)^-,0^+,(2k)^+,(2k+1)^-$ \\ \hline
\end{tabular}
\caption{Angular momentum allowed for two massless fields. See the notation in the text. 
}
\end{table}
\vskip .5cm
\end{document}